\documentclass[4p,12pt,sort&compress,preprint]{elsarticle}
\usepackage{makeidx}
\usepackage{amssymb}
\usepackage{psfrag}
\usepackage{soul}
\usepackage{amssymb,amsmath,amsthm}
\usepackage[utf8]{inputenc}
\usepackage{color}
\usepackage{graphicx}
\usepackage{amsmath,amsfonts,amsthm,amssymb,mathrsfs,bbm}
\usepackage[colorlinks=true,linkcolor=blue,citecolor=blue]{hyperref}
\usepackage[T1]{fontenc}
\usepackage{times}
\usepackage{multicol}

\journal{Chaos, Solitons and Fractals}

\begin{document}

\begin{frontmatter}
       
\title{Stable semivortex gap solitons in a spin-orbit-coupled Fermi gas}
\author[UFRO]{P. D\'{\i}az\corref{mycorrespondingauthor2}}
\cortext[mycorrespondingauthor2]{Corresponding author}
\ead{pablo.diaz@ufrontera.cl}
\author[UFRO]{H. Molinares}
\author[UTA2]{L. M. P\'{e}rez}
\author[UTA]{D. Laroze}
\author[UNAV]{J. Bragard}
\author[TAU]{B. A. Malomed}

\address[UFRO]{Departamento de Ciencias F\'{\i}sicas, Universidad de La Frontera, Casilla 54-D, Temuco 4780000, Chile.}
\address[UTA2]{Departamento de F\'{\i}sica, FACI, Universidad de Tarapac\'{a}, Casilla 7D, Arica 1000000, Chile.}
\address[UTA]{Instituto de Alta Investigaci\'{o}n, Universidad de Tarapac\'{a}, Casilla 7D, Arica 1000000, Chile.}
\address[UNAV]{Departamento de F\'{\i}sica y Matem\'{a}tica Aplicada, Universidad de Navarra, 31080, Pamplona, Spain.}
\address[TAU]{Department of Physical Electronics, School of Electrical Engineering, Faculty of Engineering, Tel Aviv University, Ramat Aviv, 69978, Israel.}

\begin{abstract}
{
We demonstrate the existence of semivortex (SV) solitons, with vorticities $0$ and $1$ in the two components, in a two-dimensional (2D) fermionic spinor system under the action of the Rashba-type spin-orbit coupling in the combination with the Zeeman splitting (ZS). In the ``heavy-atom" approximation, which was previously elaborated for the bosonic system, the usual kinetic energy is neglected, which gives rise to a linear spectrum with a bandgap. The model includes the effective Pauli self-repulsion with power $7/3$, as produced by the density-functional theory of Fermi superfluids. In the general case, the inter-component contact repulsion is included too. We construct a family of gap solitons of the SV type populating the spectral bandgap. A stability region is identified for the SV solitons, by means of systematic simulations, in the parameter plane of the cross-repulsion strength and chemical potential. The stability region agrees with the prediction of the anti-Vakhitov-Kolokolov criterion, which is a relevant necessary stability condition for systems with self-repulsive nonlinearities. We also test the stability of the SV solitons against a sudden change of the ZS strength, which initiates robust oscillations in the spin state of the soliton due to transfer of particles between the system's components.
}
\end{abstract}

\begin{keyword}
Fermi systems \sep Spin-orbit coupling \sep Semivortex solitons \sep Gap solitons \sep Soliton stability.
\end{keyword}

\end{frontmatter}


\section{Introduction}

Since the first observation of the Bose-Einstein condensates (BECs) as a
quantum state of matter in atom gases at temperatures of tens of nano-Kelvin
in 1995 \cite{Anderson1995, Davis1995,Hulet1995}, the progress in this area
of condensed-matter and atomic physics has been truly spectacular \cite%
{Bloch2008, Giorgini2008, Yukalov2009, Delibar2011, Hauke2012, Bloch2012,
Zhai2012, Zhou2013}. From the experimental point of view, the progress in
the development of magneto-optical traps, optical lattices, and the control
of interactions between particles \cite{Giorgini2008, Bloch2008,
Baumann2010, Zipkes2011, Perron2010, Best2009, Cumby2013, Deh2010, Tung2013,
Park2012, Wu2011} has led to great expansion of the variety of phenomena
observed in this realm -- in particular, due to the possibility of tuning
the effective sign and strength of inter-atomic interactions. One of fast
developing directions of the studies is the possibility to design various
forms of the synthetic (pseudo-) spin-orbit coupling (SOC) between different
atomic states, using appropriate modes of laser illumination \cite%
{Spielman2011,Galitski2013}. While in most experimental works SOC was
emulated in effectively one-dimensional (1D) BEC settings, its experimental
realization in 2D was reported too \cite{Zhan2016}.

The inclusion of SOC in the context of theoretical studies has opened the
possibility of creating stable matter-wave solitons in 2D and 3D free space,
without confining potentials \cite{Kartashov2013, Zezyulin2013,
Sakaguchi2014, Zhai2015, Sakaguchi2016a, Sakaguchi2016b, Li2017,
Luo2017,Malomed_2018,Sowinski2019, Ibrahim2021,malomed2022multidimensional}.
Furthermore, gap solitons have been predicted by considering the interplay
of SOC with the Zeeman splitting (ZS) between the components of the binary
BECs \cite{Kartashov2013, Zezyulin2013, Sakaguchi2018, Kartashov2020}.

The studies of quantum states of matter in ultracold Fermi gases have also
demonstrated great advancement. In particular, realization of SOC in atomic
Fermi systems has been reported \cite{Wang2012, Cheuk2012}. Theoretically, a
possibility of the existence of gap solitons, due to the interplay of the
Pauli self-repulsion, induced by the atomic Fermi distribution, and a
spatially-periodic potential imposed by an external optical lattice, has
been predicted in the framework of the density-functional theory \cite%
{Adhikari_2007}. The existence of 2D solitons in a free-space binary
fermionic cloud, under the action of SOC and attraction between the two
components, has been demonstrated too \cite{Diaz2019}, see Ref. \cite%
{Malomed2019} for a review.

In this work, by means of systematic numerical analysis, we report the
existence and stability of 2D gap solitons of the semivortex (SV) type,
i.e., ones with vorticities $0$ and $1$ in its two components, in the binary
Fermi system carrying SOC of the Rashba type and ZS along with the Pauli
self-repulsion produced by the density-functional approximation. To
construct the SV\ solitons, we adjust the approach, which was previously
elaborated for binary BEC\ in Ref. \cite{Sakaguchi2018}, to the fermionic
system. Namely, assuming the presence of strong SOC and ZS terms, the usual
kinetic energy is neglected, which gives rise to a bandgap in the system's
spectrum, that may be populated by solitons.

The following presentation is structured as follows: in Section \ref{SEC2},
we introduce the theoretical model and methodology. Section \ref{SEC3}
demonstrates the existence of an SV soliton family populating the system's
bandgap. In Section \ref{SEC4} we explore stability of the soliton
solutions. Section \ref{SEC5} addresses oscillatory dynamics of the solitons
initiated by a sudden change of the ZS coefficient. The paper is concluded
by Section \ref{SEC6}.

\section{The model and methods}

\label{SEC2}

In the framework of the functional-density (mean-field) theory for the Fermi
superfluid, which has been extensively elaborated in the literature \cite%
{Diaz2021, Fuchs2004, Das2003, Salasnich2000a, Salasnich2000b, Adhikari2005,
Adhikari2006a, Adhikari2007a, Adhikari2006b, Salasnich2007, Adhikari2007b},
\cite{Diaz2012, Diaz2015}, we consider the 2D spinor wave function,
\begin{equation}
\Phi (x,y,t)=(\Phi _{+}(x,y,t),\Phi _{-}(x,y,t)),  \notag  \label{spinor}
\end{equation}%
whose spatiotemporal evolution is governed by the system of coupled of
mean-field equations:
\begin{eqnarray}
i{\partial _{t}}{\Phi _{+}} &=&+{\lambda _{R}}\left( {{\partial _{x}}-i{%
\partial _{y}}}\right) {\Phi _{-}}+\Omega {\Phi _{+}}  \label{equation1} \\
&&+\left( {{{\left\vert {\Phi _{+}}\right\vert }^{4/3}}-\gamma {{\left\vert {%
\Phi _{-}}\right\vert }^{2}}}\right) {\Phi _{+}},  \notag
\end{eqnarray}%
\begin{eqnarray}
i{\partial _{t}}{\Phi _{-}} &=&-{\lambda _{R}}\left( {{\partial _{x}}+i{%
\partial _{y}}}\right) {\Phi _{+}}-\Omega {\Phi _{-}}  \label{equation2} \\
&&+\left( {{{\left\vert {\Phi _{-}}\right\vert }^{4/3}}-\gamma {{\left\vert {%
\Phi _{+}}\right\vert }^{2}}}\right) {\Phi _{-}},  \notag
\end{eqnarray}%
with terms ${{{\left\vert {\Phi _{\pm }}\right\vert }^{4/3}\Phi }}_{\pm }$
accounting for the Pauli self-repulsion in the framework of the
density-functional theory, and cubic terms $-\gamma {{{\left\vert {\Phi
_{\mp }}\right\vert }^{2}}\Phi _{\pm }}$ representing the contact
interaction between the two components (positive and negative $\gamma $
corresponds to the attractive and repulsive interactions, respectively).
Further, $\lambda _{R}$ is a real coefficient representing SOC of the Rashba
type \cite{Bychkov1984}, which has been widely studied in this form for the
binary BEC \cite{Galitski2013, Kartashov2013, Sakaguchi2014, Zhai2015,
Sakaguchi2016a}, and more recently for fermionic systems \cite{Diaz2019},
while $\Omega >0$ is the ZS strength. Note that the scaling transformation
makes it possible to fix $\lambda _{R}=\Omega \equiv 1$, which is adopted
below.

Equations (\ref{equation1}), (\ref{equation2}) are derived from the
underlying 3D system, assuming that the system is subject to strong
confinement in the transverse direction, with the trapping scale $a_{z}$.
The 2D model introduced here is appropriate for the consideration of
localized structures with lateral size $l\gg a_{z}$. Furthermore, following
Refs. \cite{Li2017} and \cite{Sakaguchi2018}, we here address the case in
which the SOC\ and ZS terms are much larger than the kinetic-energy term
(the limit of \textquotedblleft heavy atoms"), therefore the usual
Laplacians which represents the kinetic energy are dropped in Eqs. (\ref%
{equation1}) and (\ref{equation2}).

It is relevant to cast the system of equations (\ref{equation1}) and (\ref%
{equation2}) in the Hamiltonian form:%
\begin{gather}
i{\partial _{t}}{\Phi _{\pm }}=\frac{\delta H}{\delta {\Phi _{\pm }^{\ast }}}%
,  \notag \\
H=\int \int dxdy\left\{ \lambda _{R}\left[ {\Phi _{+}^{\ast }}\left( {{%
\partial _{x}}-i{\partial _{y}}}\right) {\Phi _{-}+\Phi _{+}}\left( {{%
\partial _{x}}+i{\partial _{y}}}\right) {\Phi _{-}^{\ast }}\right] \right.
\notag \\
\left. +\Omega \left( \left\vert {\Phi _{+}}\right\vert ^{2}-\left\vert {%
\Phi _{-}}\right\vert ^{2}\right) +\frac{3}{5}\left( \left\vert {\Phi _{+}}%
\right\vert ^{10/3}+\left\vert {\Phi _{-}}\right\vert ^{10/3}\right) -\gamma
\left\vert {\Phi _{+}}\right\vert ^{2}\left\vert {\Phi _{-}}\right\vert
^{2}\right\} ,  \label{H}
\end{gather}%
where $\ast $ and $\delta /\delta {\Phi _{\pm }^{\ast }}$ stand for the
complex conjugate and variational derivative, respectively. Hamiltonian $H$
and norm%
\begin{equation}
N=2\pi \int \int dxdy\left[ |\phi _{+}(x,y)|^{2}+|\phi _{-}(x,y)|^{2}\right]
\equiv N_{+}+N_{-}  \label{N}
\end{equation}%
(where $N_{\pm }$ are populations of components $\phi _{\pm }$ of the spinor
wave function), are dynamical invariants of the system of equations (\ref%
{equation1}) and (\ref{equation2}). The system also conserves the\ vectorial
momentum, which is given by the usual expression,%
\begin{equation}
\mathbf{P}=i\int \int dxdy\left( \Phi _{+}\nabla \Phi _{+}^{\ast }+\Phi
_{-}\nabla \Phi _{-}^{\ast }\right) .  \label{P}
\end{equation}

To establish the conserved angular momentum of the same system, one should
rewrite Hamiltonian (\ref{H}) for the spinor components $\Phi _{-}$ and $%
\tilde{\Phi}_{+}\equiv \exp \left( -i\theta \right) \Phi _{+}$ in polar
coordinates $\left( r,\theta \right) $:%
\begin{gather}
H=\int_{0}^{\infty }rdr\int_{0}^{2\pi }d\theta \left\{ \lambda _{R}\left[
\tilde{\Phi}{_{+}^{\ast }}\left( \frac{\partial }{\partial r}{-}\frac{{i}}{r}%
\frac{\partial }{\partial \theta }\right) {\Phi _{-}+\tilde{\Phi}_{+}}\left(
\frac{\partial }{\partial r}{-}\frac{{i}}{r}\frac{\partial }{\partial \theta
}\right) {\Phi _{-}^{\ast }}\right] \right.   \notag \\
\left. +\Omega \left( \left\vert \tilde{\Phi}{_{+}}\right\vert
^{2}-\left\vert {\Phi _{-}}\right\vert ^{2}\right) +\frac{3}{5}\left(
\left\vert \tilde{\Phi}{_{+}}\right\vert ^{10/3}+\left\vert {\Phi _{-}}%
\right\vert ^{10/3}\right) -\gamma \left\vert \tilde{\Phi}{_{+}}\right\vert
^{2}\left\vert {\Phi _{-}}\right\vert ^{2}\right\} .  \label{H2}
\end{gather}%
The invariance of this expression with respect to the arbitrary rotation in
the $\left( x,y\right) $ plane, $\theta \rightarrow \theta +\Delta \theta $,
implies the conservation of the respective angular momentum, as defined by
the Noether theorem \cite{Bogoliubov1973}. It can be written, eventually, in
terms of the original components ${\Phi _{\pm }}$:
\begin{gather}
M=i\int_{0}^{\infty }rdr\int_{0}^{2\pi }d\theta \left( {\Phi _{-}^{\ast }%
\frac{\partial }{\partial \theta }\Phi _{-}+\tilde{\Phi}_{+}^{\ast }\frac{%
\partial }{\partial \theta }\tilde{\Phi}_{+}}\right)   \notag \\
\equiv \int_{0}^{\infty }rdr\int_{0}^{2\pi }d\theta \left[ i\left( {\Phi
_{-}^{\ast }\frac{\partial }{\partial \theta }\Phi _{-}+\Phi _{+}^{\ast }%
\frac{\partial }{\partial \theta }\Phi _{+}}\right) {+}\left\vert \Phi
_{+}\right\vert ^{2}\right] .  \label{M}
\end{gather}%
We have checked that all numerical simulations reported below conserve the
total norm and angular momentum of the spinor system.

\begin{figure}[h]
\centering
\includegraphics[width=0.7\linewidth]{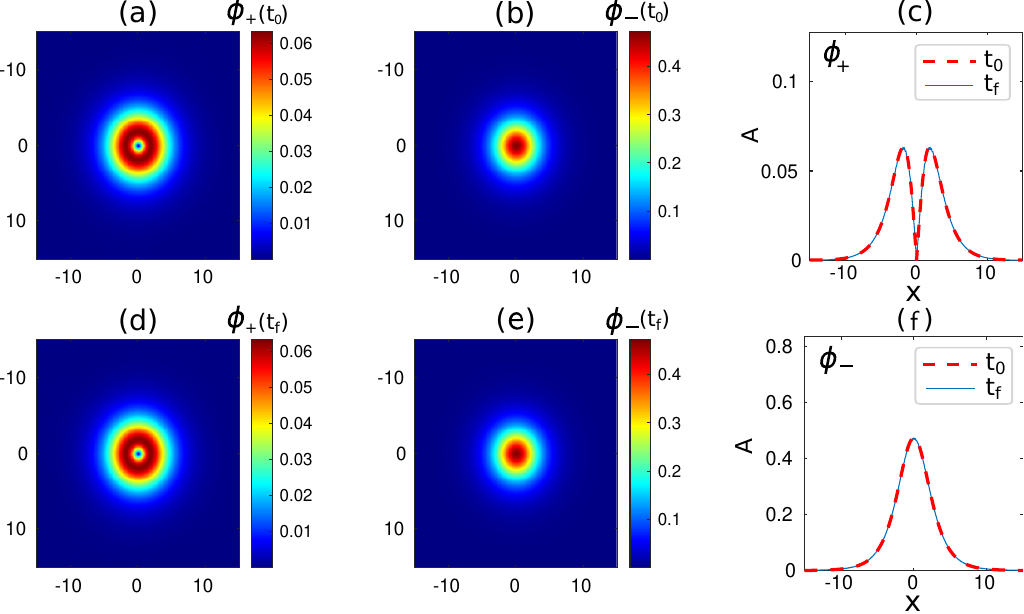}
\caption{Contour plots of the vortical and fundamental (zero-vorticity)
components of the stable SV soliton, $\left\vert \Phi _{+}(x,y)\right\vert $
and $\left\vert \Phi _{-}(x,y)\right\vert $, respectively, in the $(x,y)$
plane. (a,b) The input (at $\mathrm{t}_{0}=0$), produced by the numerical
solution of Eqs. (\protect\ref{radial_1}) and (\protect\ref{radial_2}).
(d,e) The result produced by the evolution of the input (simulations of Eqs.
(\protect\ref{equation1}) and (\protect\ref{equation2})) at $\mathrm{t}_{%
\mathrm{f}}=1000$. To verify the soliton's stability, the initial and final
shapes of cross-sections of its vortical and fundamental components are
compared, respectively. in panels (c) and (f). Parameters are $\protect%
\gamma =0$ (no nonlinear interaction between the components), $\protect%
\lambda _{R}=1$, $\Omega =1$, $m_{+}=-1$, $m_{-}=0$, and $\protect\mu =-0.88$%
. Note that this value of $\protect\mu $ belongs to the corresponding
bandgap (\protect\ref{gap}).}
\label{fig3}
\end{figure}

To define SV soliton solutions, we follow the lines of the analysis
developed in the context of the mean-field theory for the spin-orbit-coupled
BEC in Ref. \cite{Sakaguchi2018}. To this end, we introduce an ansatz based
on the separation of variables in the polar coordinates $\left( r,\theta
\right) $, with integer vorticities
\begin{equation}
m_{-},m_{+}=m_{-}-1  \label{mm}
\end{equation}%
of the two components of the spinor wave function:
\begin{equation}
{\Phi _{\pm }(r,\theta ,t)}={e^{-i\mu t}}{e^{i{m_{\pm }}\theta }}{\phi _{\pm
}(r)},  \label{equation3}
\end{equation}%
where real $\mu $ is the chemical potential, and $\phi _{\pm }(r)$ are real
functions. The relation between $m_{+}$ and $m_{-}$ adopted in Eq. (\ref{mm}%
) is imposed by the form of the Rashba SOC terms in Eqs. (\ref{equation1})
and (\ref{equation2}), cf. Ref. \cite{Sakaguchi2014}. Accordingly, the
angular momentum (\ref{M}) for ansatz (\ref{equation3}) is $M_{m_{-}}=\left(
1-m_{-}\right) N_{-}+\left( 2-m_{-}\right) N_{+}$.

Substituting the ansatz, defined by Eqs. (\ref{mm}) and (\ref{equation3}),
in Eqs. (\ref{equation1}) and (\ref{equation2}), we arrive at the following
equations for the radial functions:
\begin{eqnarray}
\frac{d}{{dr}}{\phi _{-}} &=&\mu {\phi _{+}}-{\phi _{+}}-\frac{{m_{-}}}{r}{%
\phi _{-}}  \label{general_radial_1} \\
&&-\left( {{{{\phi _{+}^{4/3}}}}-\gamma {{{\phi _{-}^{2}}}}}\right) {\phi
_{+}},  \notag
\end{eqnarray}%
\begin{eqnarray}
\frac{d}{{dr}}{\phi _{+}} &=&-\mu {\phi _{-}}-{\phi _{-}}+\frac{{m_{+}}}{r}{%
\phi _{+}}  \label{general_radial_2} \\
&&+\left( {{{{\phi _{-}^{4/3}}}}-\gamma {{{\phi _{+}^{2}}}}}\right) {\phi
_{-}},  \notag
\end{eqnarray}%
where, as said above, $\lambda _{R}=\Omega =1$ is fixed by means of scaling,
the remaining free parameters being $\gamma $ and $\mu $. The family of SV
solitons is characterized by the dependence of norm (\ref{N}) on $\mu $, see
below.

Following Ref. \cite{Sakaguchi2018}, we are interested in finding SV\ gap
solitons in the fundamental (ground) state, which corresponds to $m_{-}=0$
or $m_{+}=0$, while excited states, with $m_{-}\cdot m_{+}\neq 0$, are
expected to be unstable. Thus, setting $m_{-}=0$ and $m_{+}=-1$, in
agreement with Eq. (\ref{mm}), Eqs. (\ref{general_radial_1}) and (\ref%
{general_radial_2}) reduce to%
\begin{eqnarray}
\frac{d}{{dr}}{\phi _{-}} &=&{\mu }{\phi _{+}}-{\phi _{+}}  \label{radial_1}
\\
&&-\left( {{{{\phi _{+}^{4/3}}}}-\gamma {{{\phi _{-}^{2}}}}}\right) {\phi
_{+}},  \notag
\end{eqnarray}%
\begin{eqnarray}
\frac{d}{{dr}}{\phi _{+}} &=&-{\mu }{\phi _{-}}-{\phi _{-}}-\frac{{\lambda
_{R}}}{r}{\phi _{+}}  \label{radial_2} \\
&&+\left( {{{{\phi _{-}^{4/3}}}}-\gamma {{{\phi _{+}^{2}}}}}\right) {\phi
_{-}}.  \notag
\end{eqnarray}%
An alternative option is to set $m_{+}=0$, $m_{-}=+1$, and $\Omega =-1$,
which corresponds to the ground-state SV which is a mirror image of the one
corresponding to Eqs. (\ref{radial_1}) and (\ref{radial_2}).

For exponentially localized soliton solutions, Eqs. (\ref{radial_1}) and (%
\ref{radial_2}) take the linearized asymptotic form of at $r\rightarrow
\infty .$ It is easy to see that the respective system of two linear
first-order equations can be reduced to a single equation for $\phi _{+}(r)$%
, which is tantamount to the equation for the Bessel functions:%
\begin{equation}
\frac{d^{2}{\phi _{+}}}{{dr}^{2}}+\frac{1}{r}\frac{d{\phi _{+}}}{{dr}}%
-\left( \frac{\Omega ^{2}-\mu ^{2}}{\lambda _{R}^{2}}+\frac{1}{r^{2}}\right)
{\phi _{+}=0,}  \label{second}
\end{equation}%
the asymptotic expression for $\phi _{-}(r)$ being%
\begin{equation}
\phi _{-}(r)=-\frac{\lambda _{R}}{\Omega +\mu }\left( \frac{d}{{dr}}+\frac{1%
}{r}\right) {\phi _{+}(r).}  \label{phi-}
\end{equation}%
As it follows from Eq. (\ref{second}), the localized solutions may exist in
the spectral \emph{bandgap}, which is essentially the same as in the similar
BEC system \cite{Sakaguchi2018},%
\begin{equation}
-\Omega <\mu <+\Omega .  \label{gap}
\end{equation}%
Accordingly, the localized modes populating the bandgap are called gap
solitons, as said above. At $r\rightarrow \infty $, their asymptotic form is
given by the appropriate solution of Eq. (\ref{second}),%
\begin{equation}
\phi _{+}(r)=\phi _{0}^{(+)}K_{1}\left( \frac{\sqrt{\Omega ^{2}-\mu ^{2}}}{%
\lambda }r\right) ,  \label{K}
\end{equation}%
where $K_{1}$ is the modified Bessel function of the second kind, which
exponentially decays $\sim r^{-1/2}\exp \left( -\frac{\sqrt{\Omega ^{2}-\mu
^{2}}}{\lambda }r\right) $, and $\phi _{0}^{(+)}$ is an arbitrary constant.

An asymptotic expansion of the relevant solution to Eqs. (\ref{radial_1})
and (\ref{radial_2}) can also be constructed at $r\rightarrow 0$:%
\begin{eqnarray}
\phi _{+}(r) &=&-\frac{\Omega +\mu }{\lambda _{R}}\phi _{0}^{(-)}\left( 1-%
\frac{\left( \phi _{0}^{(-)}\right) ^{4/3}}{\Omega +\mu }\right) r+\mathcal{O%
}\left( r^{3}\right) ,  \notag \\
&&  \label{r-->0} \\
\phi _{-}(r) &=&\phi _{0}^{(-)}+\frac{\Omega ^{2}-\mu ^{2}}{2\lambda ^{2}}%
\phi _{0}^{(-)}\left( 1-\frac{\left( \phi _{0}^{(-)}\right) ^{4/3}}{\Omega
+\mu }\right) r^{2}+\mathcal{O}\left( r^{4}\right) ,  \notag
\end{eqnarray}%
where $\phi _{0}^{(-)}$ is another arbitrary constant.

Before producing systematic results for families of SV solitons, we display
a generic example of a stable one in Fig. \ref{fig3}. Note that the chemical
potential corresponding to this solution, $\mu =-0.88$, indeed belongs to
bandgap (\ref{gap}).

\section{Semivortex gap solitons}

\label{SEC3}

In this section, we summarize results for the SV gap solitons produced by
the numerical solution of radial equations (\ref{radial_1}) and (\ref%
{radial_2}). They were solved in the region of $0\leq r\leq 30$, with the
spatial mesh size $\Delta r=10^{-5}$, by means of the shooting method
together with the Euler method employed for the integration \cite{Shooting}.
The boundary conditions at $r=0$ were taken as $\phi _{+}(r=0)=0$ and
\begin{equation}
\phi _{-}(r=0)=A>0.  \label{A}
\end{equation}%
In the framework of this scheme, we looked for a value of $A$ in Eq. (\ref{A}%
) for which both fields tend to zero at $r\rightarrow \infty $ (in fact,
this means $\phi _{\pm }(r=30)=0$ in the present setting), to build
localized solutions. Several localized solutions, with different values of
the norm (\ref{N}), were thus found for increasing values of $A$. We
selected the solution with the lowest value of $N$, which corresponds to the
lowest $A$, as fundamental soliton. Higher-order ones, with greater $\ A$
and greater $N$, correspond to excited localized states, which are expected
to be completely unstable, as suggested by Ref. \cite{Sakaguchi2018}, where
this conclusion was made in the context of the binary (spin-orbit-coupled)
BEC model.
\begin{figure}[h]
\centering
\includegraphics[width=0.26\linewidth]{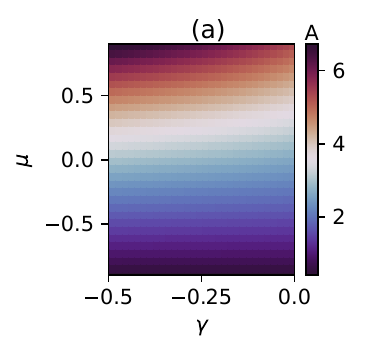} \hspace{-0.6cm} %
\includegraphics[width=0.268\linewidth]{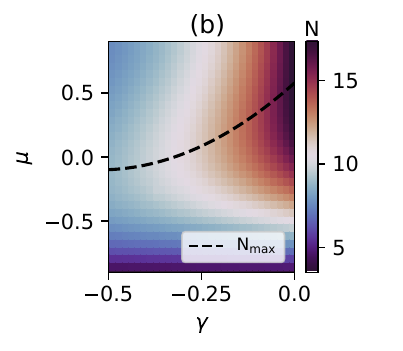} \hspace{-0.6cm} %
\includegraphics[width=0.28\linewidth]{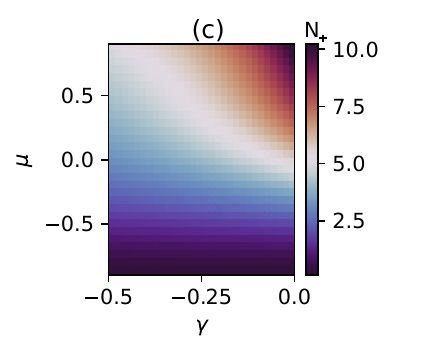} \hspace{-0.6cm} %
\includegraphics[width=0.268\linewidth]{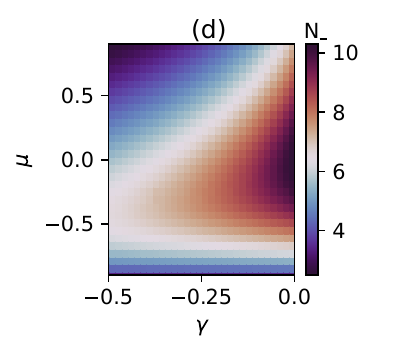}
\caption{(a) The heat map for amplitude (\protect\ref{A}) in the plane of
the chemical potential $\protect\mu $ and cross-interaction\ coefficient $%
\protect\gamma $. Panels (b) and (c,d) display the same for the total norm $%
N $ and populations of the two components, see Eq. (\protect\ref{N}). }
\label{fig1}
\end{figure}

Figure \ref{fig1} summarizes our findings by means of heat maps exhibiting
values of\ the amplitude (\ref{A}), along with the total norm $N$ and
component populations $N_{\pm }$ (see Eq. (\ref{N})), in the plane of free
parameters $\left( \gamma ,\mu \right) $ of the system of Eqs. (\ref%
{radial_1}) and (\ref{radial_2}), in the ranges of $\gamma \in \lbrack
-0.5,0]$ and $\mu \in \lbrack -0.9,+0.9]$. We display the results only for $%
\gamma \leq 0$, as the case of $\gamma >0$, i.e., attraction between the
components of the spinor wave function, leads to instability driven by the
collapse. Indeed, a simple estimate following the lines of the general
collapse theory \cite{Berge1998} demonstrates that, in the 2D system with
the dispersion represented by the first-order spatial derivatives, the
critical or supercritical collapse occurs, respectively, under the action of
a quadratic attractive nonlinearity, or any nonlinearity stronger than
quadratic, including the cubic terms corresponding to $\gamma >0$ in Eqs. (%
\ref{equation1}) and (\ref{equation2}).

Figure \ref{fig1}(a) demonstrates that amplitude $A$ notably increases with
the increase of chemical potential $\mu $, slightly decreasing with the
increase of the negative cross-interaction coefficient $\gamma $. In panel
(b),\ the total norm $N$ increases with the increase of $\mu $ at a fixed
value of $\gamma $, until it reaches a maximum value at points belonging to
the curve labeled $N_{\max }$ in panel (b). The location of this line plays
a key role in the discussion of the soliton stability below. Naturally, the
maximum value of $N$ increases for a less repulsive cross-interaction, i.e.,
smaller values of $|\gamma |$), allowing for the existence of more massive
solitons. Panels (c) and (d) demonstrate that the zero-vorticity spinor
component ($\psi _{-}$) dominates at $\mu <0$, while at $\mu >0$ the norms
of both components take close values, especially for smaller values of $%
|\gamma |$.

\begin{figure}[h]
\centering
\includegraphics[width=0.27\linewidth]{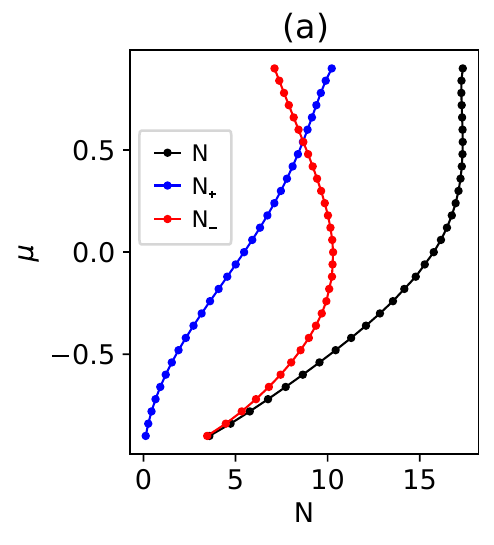} \hspace{0.3cm} %
\includegraphics[width=0.258\linewidth]{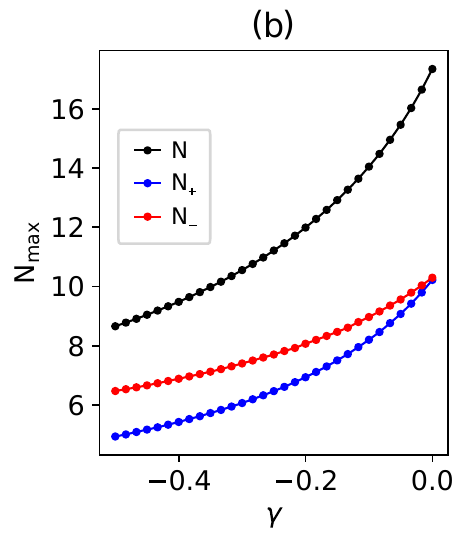} \hspace{0.3cm} %
\includegraphics[width=0.31\linewidth]{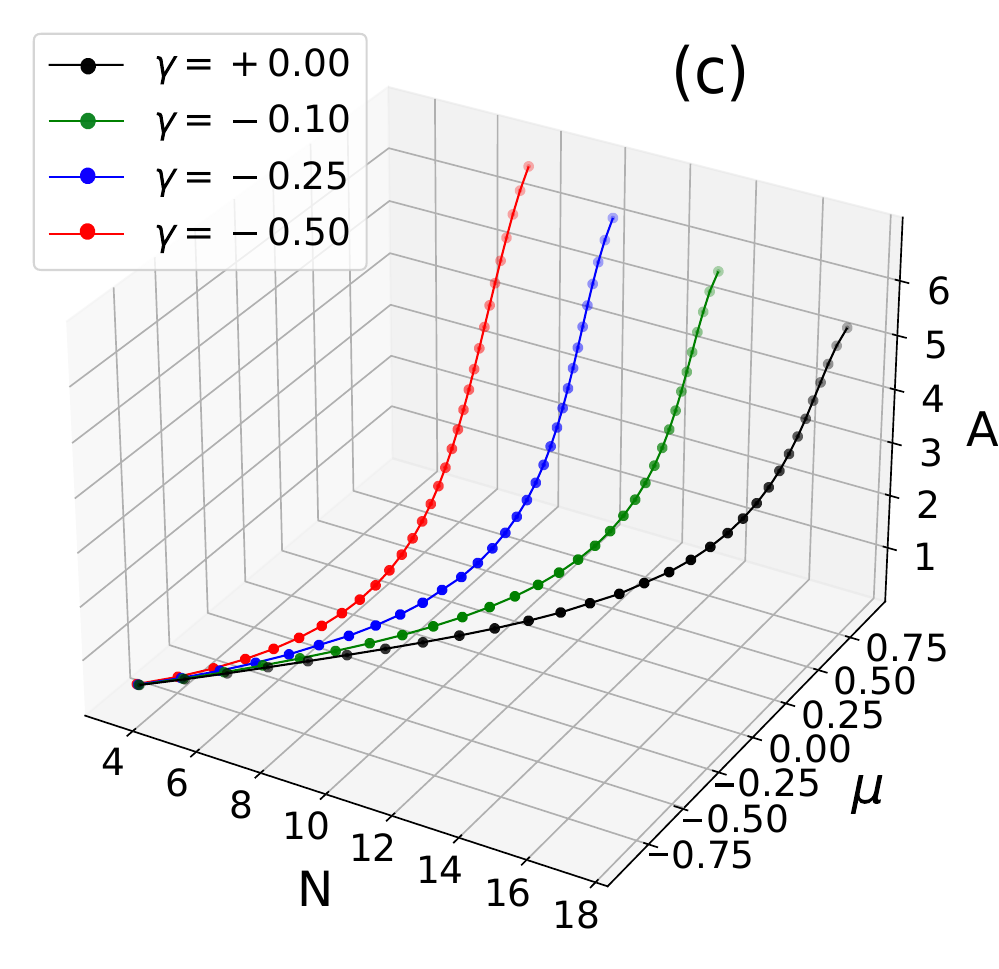}
\caption{(a) Curves $\protect\mu (N)$, $\protect\mu (N_{+})$ and $\protect%
\mu (N_{-})$ for the family of SV solitons at $\protect\gamma =0$. (b) The
largest values of the norm total and component norms from panels (b)-(d) of
Fig. \protect\ref{fig1}. (c) Amplitude $A$ (see Eq. (\protect\ref{A})) vs.
the total norm $N$ and chemical potential $\protect\mu $ for different
values of the cross-repulsion coefficient $\protect\gamma $.}
\label{fig2}
\end{figure}

\begin{figure}[h!]
\centering
\includegraphics[width=0.34\linewidth]{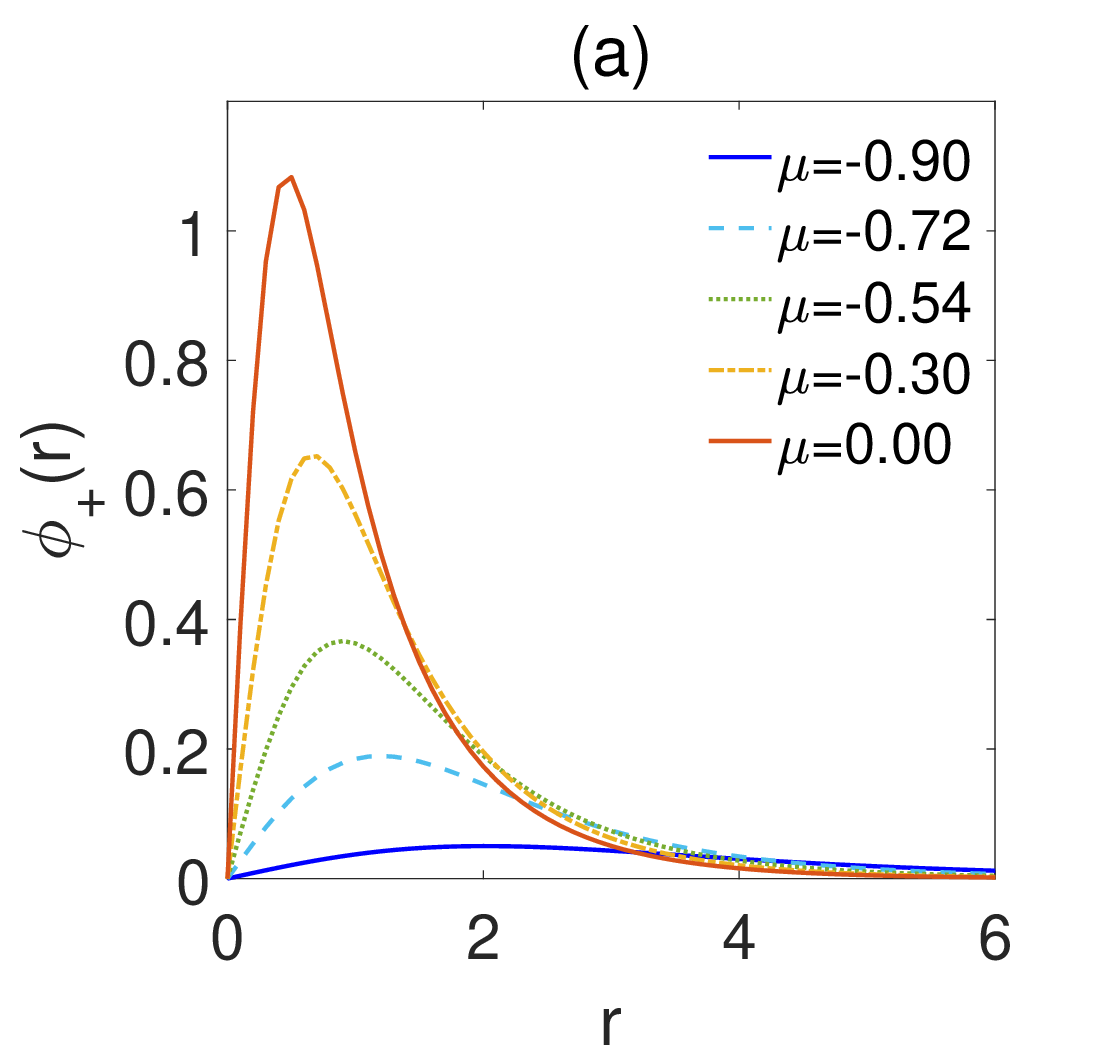} %
\includegraphics[width=0.34\linewidth]{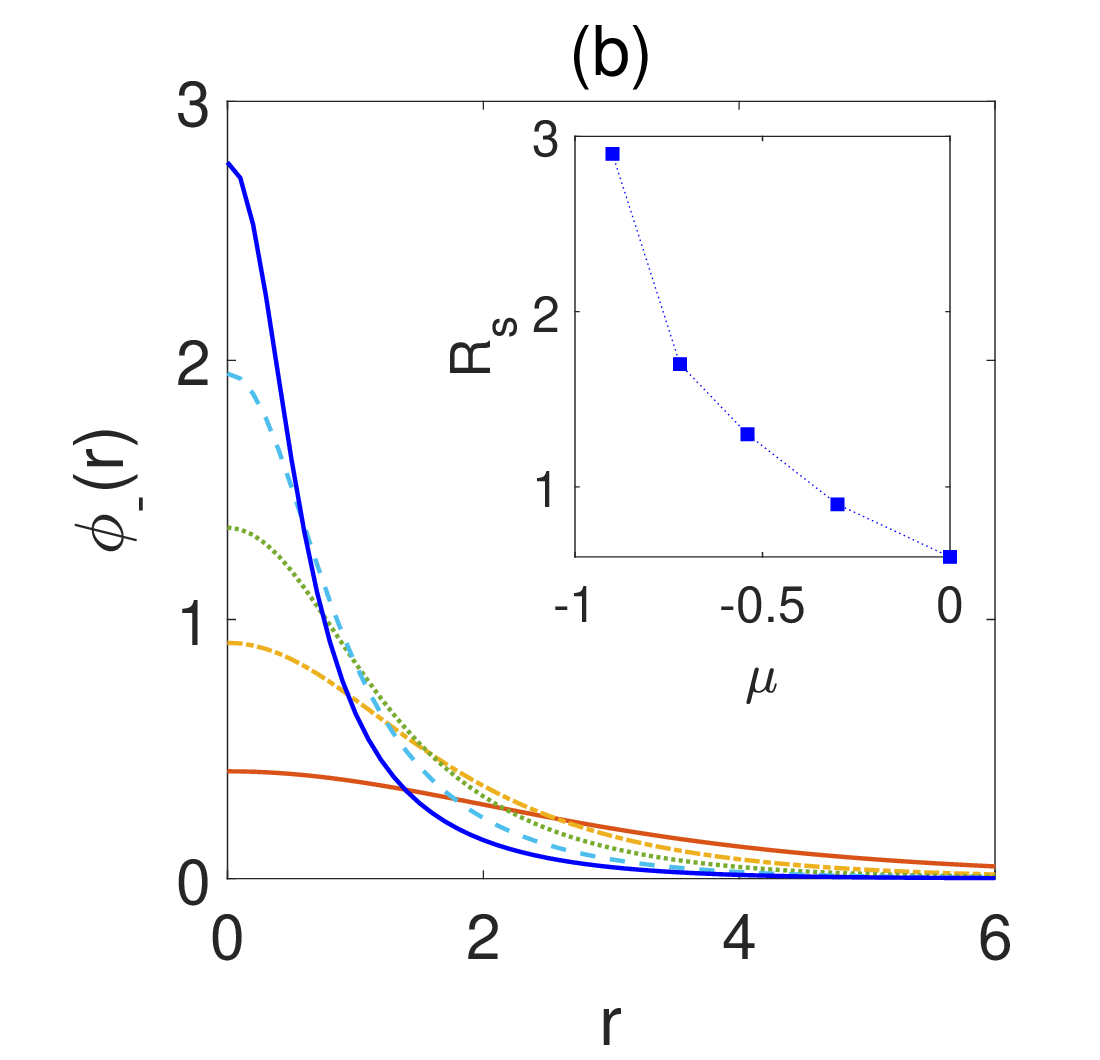}
\caption{Radial profiles of the vortical component $\protect\phi _{+}$ (a)
and zero-vorticity one $\protect\phi _{-}$ (b) for fixed $\protect\gamma %
=-0.25$.}
\label{Newfig}
\end{figure}

For further visualization of the findings, in Fig. \ref{fig2}(a) we present
the relation between the chemical potential $\mu $ and the total norm $N$,
along with the populations of the two components, for $\gamma =0$. In
particular, the two populations are equal for $\mu \approx 0.5$. The
population distribution minimizing the Zeeman energy, i.e., having $%
N_{-}>N_{+}$ and thus helping to stabilize the SV solitons, is chosen by the
system at $\mu <0.5$. The distribution is inverted at $\mu >0.5$, leading to
the increase of the Zeeman energy and thus destabilizing the solitons, as
shown below.

As demonstrated above in Fig. \ref{fig1}(b), the total norm attains a
maximum value, $N_{\max }$, at a particular value of $\mu $. The dependence
of $N_{\max }$ on the inter-component repulsion coefficient $\gamma $ is
plotted in Fig. \ref{fig2}(b). This figure shows that, naturally, $N_{\max }$
decreases as the cross repulsion becomes stronger. Additionally, in this
range of values of $\gamma $, the norm is predominantly concentrated in the
zero-vorticity spinor without vorticity ($\Phi _{-}$) which, as said above,
helps to stabilize the SV solitons by reducing its Zeeman energy. Further,
the dependence of the soliton's amplitude $A$ on the norm $N$ and,
simultaneously, on the chemical potential $\mu $ (recall that $\mu $ is not
an independent variable, but a function of $N$, pursuant to Fig. \ref{fig2}%
(c)) is plotted in Fig. \ref{fig2}(c), for several different values of $%
\gamma $.

Radial profiles of the vortical and zero-vorticity components of the SV
solitons, $\phi _{+}$ and $\phi _{-}$ (see Eq. (\ref{equation3})), are
presented in Fig. \ref{Newfig} for five different values of $\mu $ and fixed
$\gamma =-0.25$. We observe that, in agreement with Fig. \ref{fig1}(a), the
solitons increase their amplitude and become narrower with the increase of $%
\mu $ towards positive values. The inset in Fig. \ref{Newfig}(b) shows the
effective radius $R_{s}$ of the soliton, which we define as the values of $r$
at which the local value of the zero-vorticity component, $\phi _{-}(r)$,
falls to $0.5$ of $A\equiv \phi _{-}(r=0)$ (see Eq. (\textit{\ref{A}})).
These results demonstrate that the matter density in the solitons steeply
increases with the chemical potential. As shown in the next section, this
trend leads to the onset of instability of the SV solitons.

\section{Stability of the semivortex solitons.}

\label{SEC4}

The above analysis suggests that variation of the cross-repulsion
coefficient $\gamma $ impacts on characteristics and shapes of the SV
solitons, therefore one may expect that it also affects their stability. To
address this issue, we have studied the stability of the soliton solutions
in the parametric space $\left( \gamma ,\mu \right) $ by means of direct
simulation of Eqs. (\ref{equation1}) and (\ref{equation2}), taking, as the
input, the ansatz (\ref{equation3}) with the radial wave functions produced
by the numerical solution of Eqs. (\ref{radial_1}) and (\ref{radial_2}). The
simulations were performed by means of the fourth-order Runge-Kutta $4$
method, with a fixed time-marching step $\Delta t=10^{-4}$. Spatial
derivatives were approximated using the standard centered second-order
finite-difference scheme, with mesh sizes $\Delta x=\Delta y=0.025$.

\begin{figure}[h]
\centering
\includegraphics[width=0.74\linewidth]{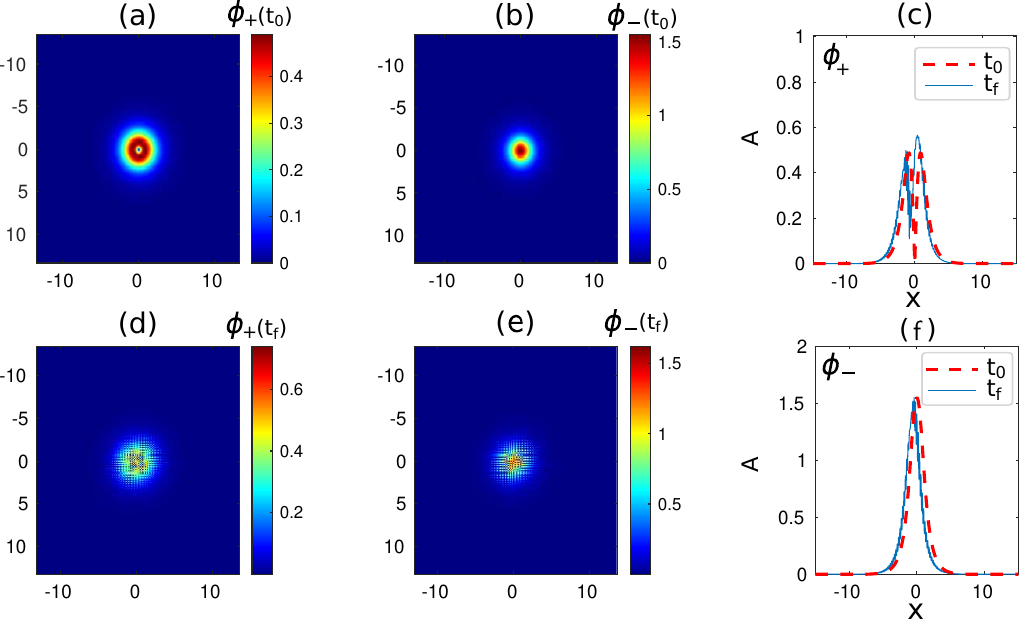}
\caption{An example of the unstable evolution of an SV soiliton with $%
\protect\gamma =0$ and $\protect\mu =-0.44$, produced by simulations of Eqs.
(\protect\ref{equation1}) and (\protect\ref{equation2}). (a,b) Contour plots
of the vortical and zero-vorticity components, $|\protect\phi _{+}(r)|$ and $%
|\protect\phi _{-}(r)|$, in the input. (d,e) The same at titme $t_{\mathrm{f}%
}=160$. (c,f) Comparison of the initial and final profiles of the two
components.}
\label{fig4}
\end{figure}

An example of an unstable SV soliton is presented in Fig. \ref{fig4} for
parameters $\gamma =0$ and $\mu =-0.44$ (different from the example of the
stable soliton presented above in Fig. \ref{fig3}, which pertains to $\mu
=-0.88$). As seen in Figs. \ref{fig4}(d) and (e), the soliton's instability
is evident at time $t=160$. Panels (c) and (f) display the comparison of the
initial soliton's profiles and ones produced by the evolution at $t=160$
(blue and red curves, respectively). The examples presented in \ Figs. \ref%
{fig3} and \ref{fig4} reveal that the stability of the solitons strongly
depends on values of the parameters, such as the chemical potential $%
\mu
$.

\begin{figure}[h]
\centering
\includegraphics[width=0.31\linewidth]{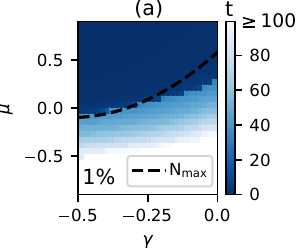} \includegraphics[width=0.31%
\linewidth]{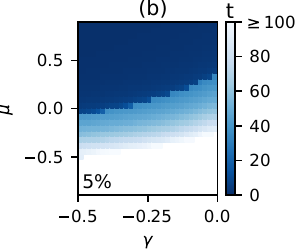} \includegraphics[width=0.264\linewidth]{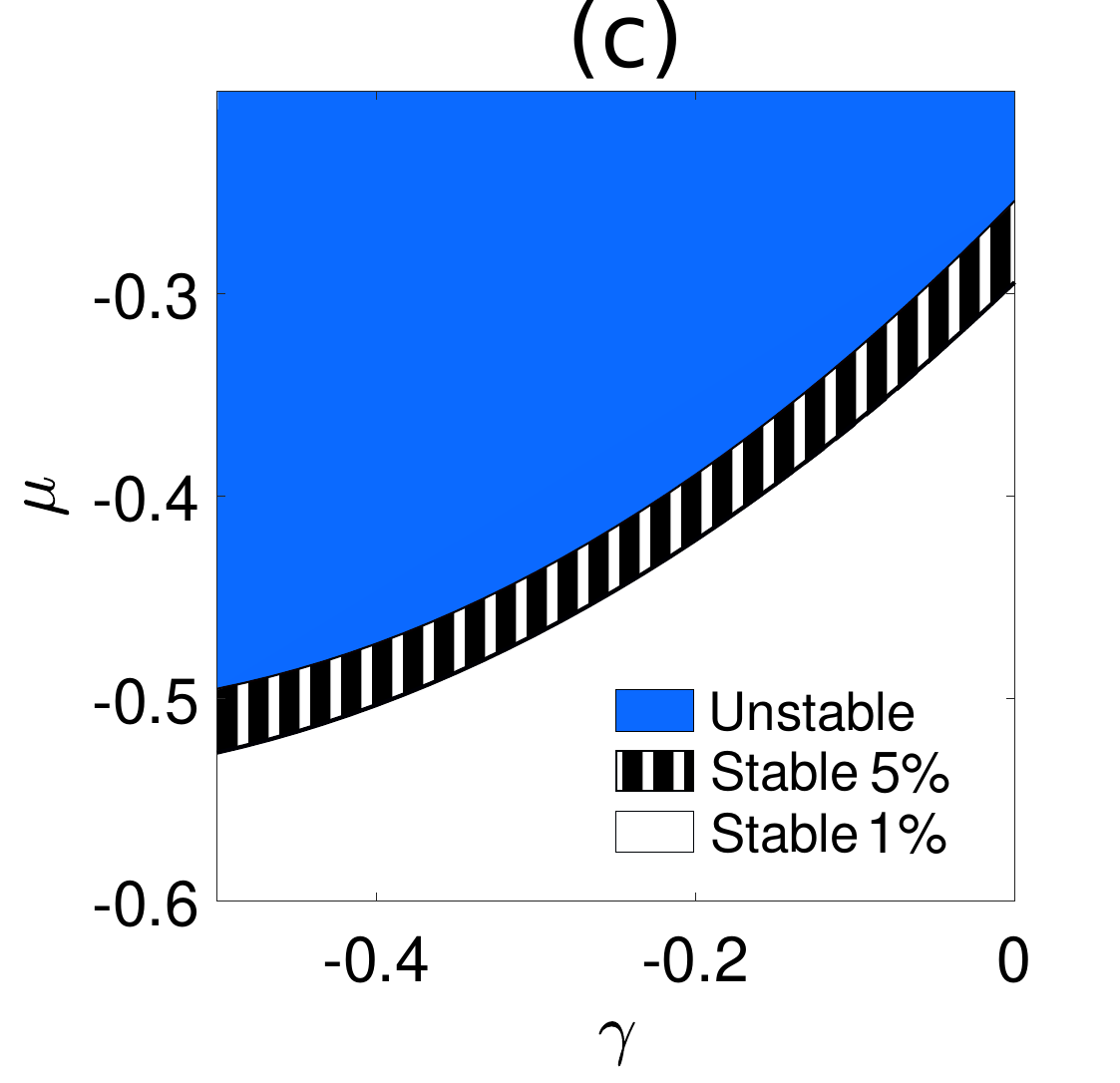}
\caption{The lifetime of the SV solitons as a function of the chemical
potential $\protect\mu $ and the cross-repulsion coefficient $\protect\gamma
$. Panels (a) and (b) display heat maps of the time at which the amplitude
variation reaches the levels of $1\%$ and $5\%$, respectively. (c) A sharper
visualization of the regions where the same variations occur at $t\geq 100$
(for the stable solitons at sufficiently negative $\protect\mu $, the
variations never take place).}
\label{stability}
\end{figure}

In the case of a dominant attractive nonlinearity, the necessary stability
condition for solitons is given by the well-known Vakhitov-Kolokolov (VK)
criterion, $d\mu /dN<0$ \cite{Vakhitov1973,Berge1998,Sakaguchi2018}. In the
present case, the Pauli repulsion term provides the dominant nonlinearity.
As mentioned above, Eqs. (\ref{radial_1}) and (\ref{radial_2}) are similar
to those addressed in the context of the BEC gap-soliton model in Ref. \cite%
{Sakaguchi2018}, in which, however, the interaction was attractive. It is
known that, in the case of the dominant repulsive nonlinearity, the VK
stability criterion is replaced by the \textit{anti-VK} one, \textit{viz}., $%
d\mu /dN>0$ \cite{SakaguchiVK, Dong2023}.

The stability of the SV-soliton family was tested by dint of the following
procedure: the evolution of the amplitude at the center of the soliton, $%
|\Phi _{-}(x=y=0,t)|$, was monitored, to record the time at which the
amplitude deviates by $1\%$ and $5\%$ from the initial value, as shown in
Figs. \ref{stability}(a) and (b), respectively. Note that there is no
significant difference between the figures. This means that, once the system
is destabilized, attaining the amplitude variation of $1\%$, the variation
of $5\%$ was reached soon afterwards. The small difference between the times
corresponding to $1\%$ and $5\%$ suggests that the $1\%$ criterion is a
sound indicator of the soliton stability. Next, we note that, for all values
of $\gamma $, there is a value of $\mu $ for which the instability time is
vanishingly small (indicating strong instability in the darkest regions of
the maps). This critical value of $\mu $ becomes larger as $\gamma $
approaches zero. Something similar happens in the opposite limit,\ which
corresponds to the transition to stability: when the time necessary for the
onset of the relative variation in the size of $1\%$ exceeds $t=100$ (the
lighter region of the maps).

Figure \ref{stability}(c) displays the effective stability boundaries,
displaying the lines below which the solitons remain stable according to the
$5\%$ and $1\%$ criteria at $t\geq 100$. Returning to the parameter area in
which the soliton's survival time tends to zero (the dark zone in Figs. \ref%
{stability}(a,b)), it is seen, comparing it with Fig. \ref{fig1}(b), that
the edge of this area coincides, approximately, with the curve at which $N$
attains the maximum, as a function of $\mu $, for each fixed value of $%
\gamma $ (as indicated by the dashed line in Fig. \ref{fig1}(b)).

\begin{figure}[h]
\centering
\includegraphics[width=0.32\linewidth]{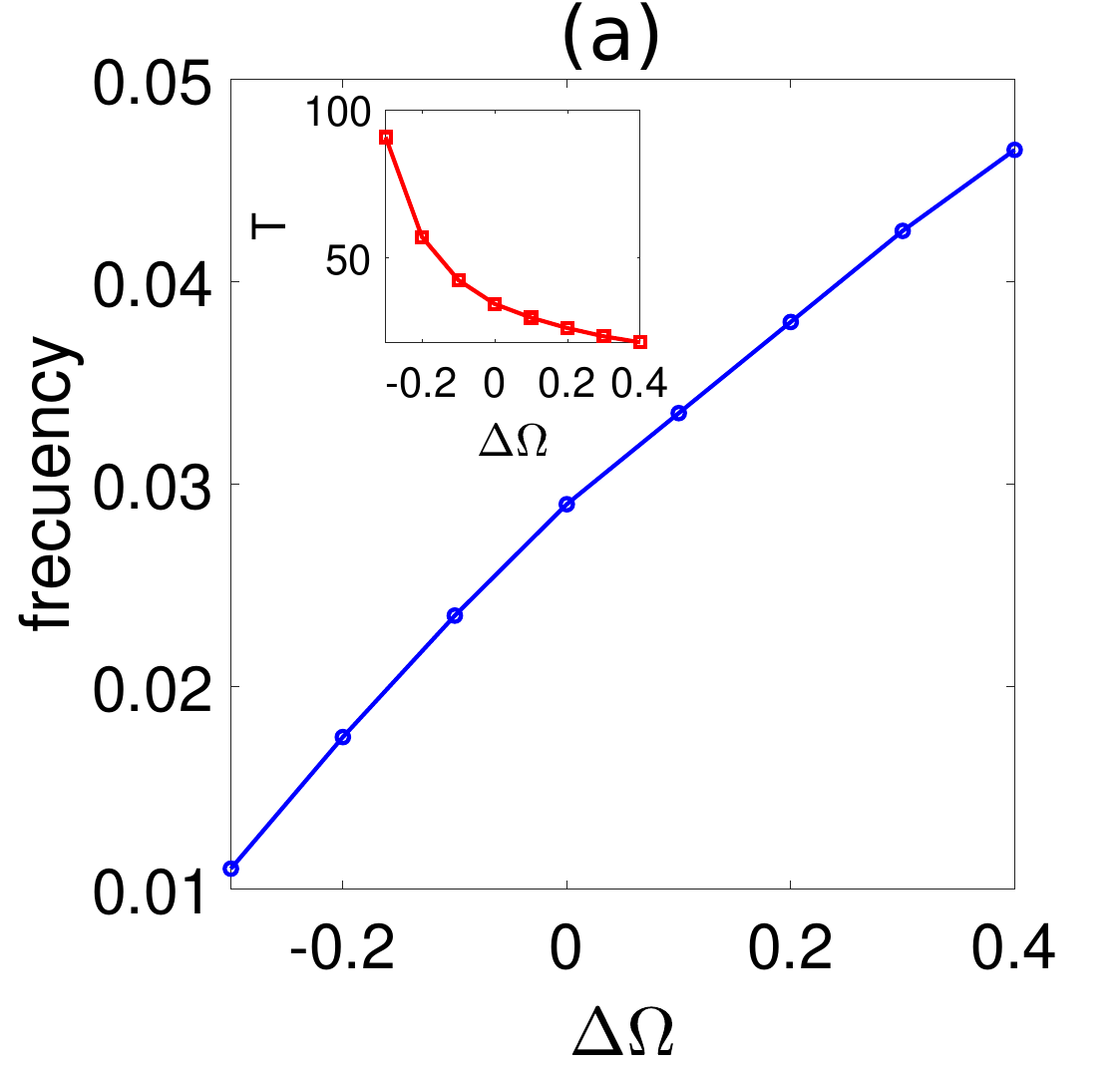} \includegraphics[width=0.32%
\linewidth]{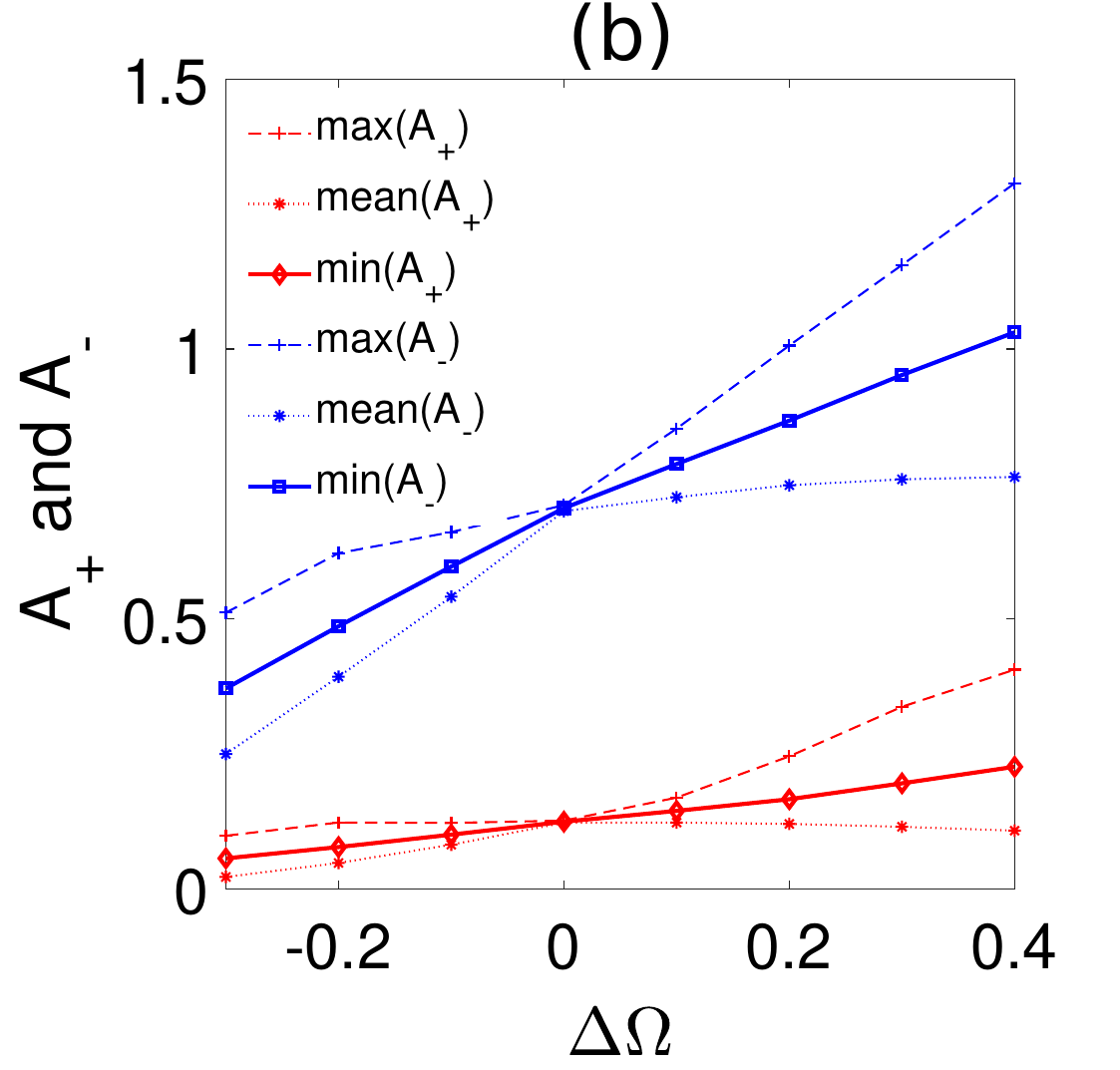} \includegraphics[width=0.32\linewidth]{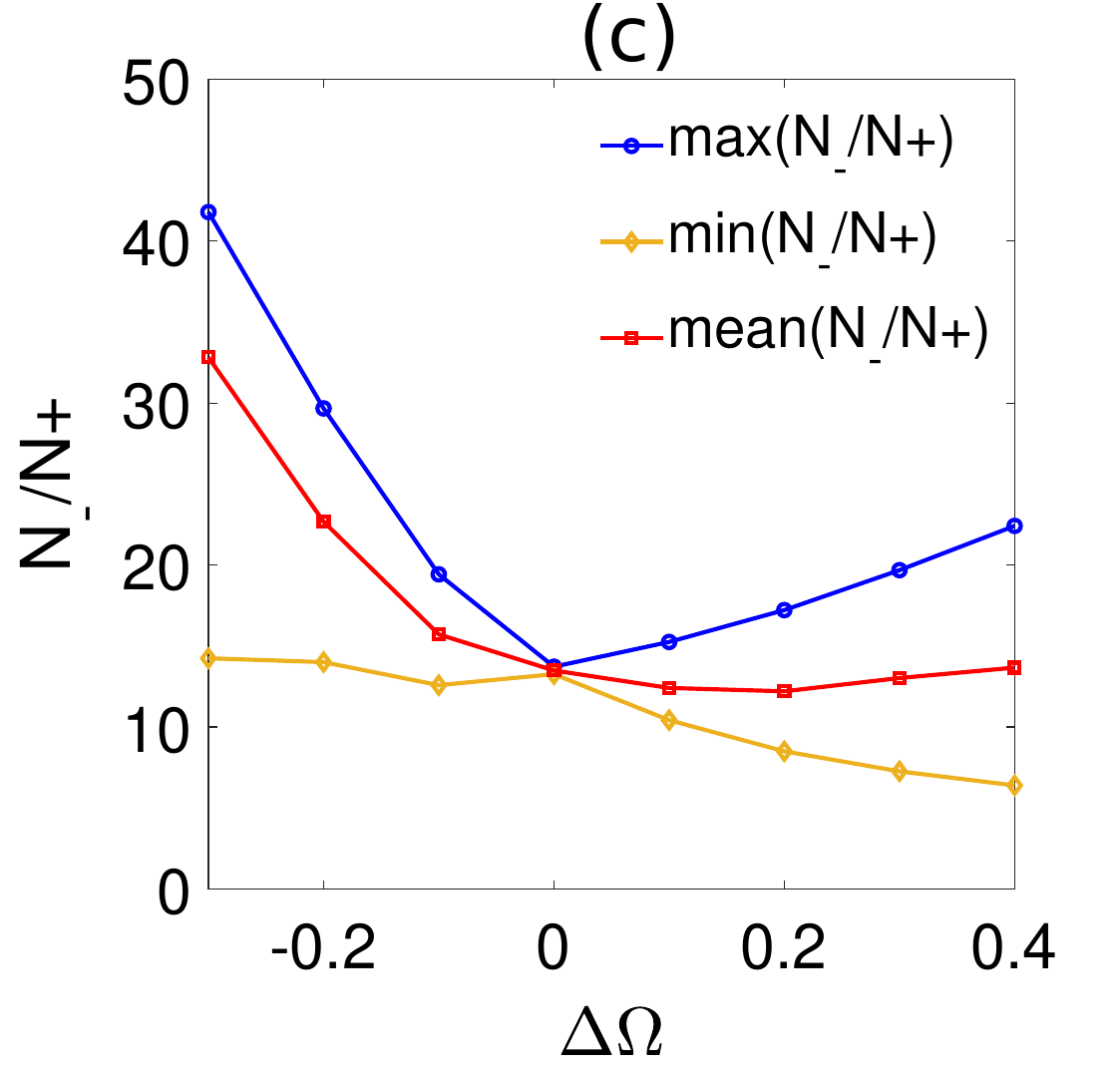}
\caption{(a) The frequency of the internal oscillations of the soliton,
excited by the sudden change $\Delta \Omega $ of the ZS strength, vs. $%
\Delta \Omega $. (b) Amplitudes $A_{+}(t)$ and $A_{-}(t)$ as functions of $%
\Delta \Omega $. (c) The relative weight of the spinor components, $%
N_{-}/N_{+}(t)$, as a function of $\Delta \Omega $. Other parameters are $%
\protect\gamma =0$, $\protect\mu =-0.8$, $\protect\lambda _{R}=1$, $\Omega
=1 $, $m_{+}=-1$, $m_{-}=0$.}
\label{fig6}
\end{figure}

Note that, as Fig. \ref{fig2}(a) demonstrates (at least, for $\gamma =0$),
the anti-VK criterion, $d\mu /dN>0$, definitely holds in the stability areas
which are identified in Fig. \ref{stability}, while the instability is
correlated with the region in which this criterion becomes indefinite
(corresponding to $dN/d\mu \approx 0$). Furthermore, if the same criterion
is (tentatively) applied to the norm $N_{-}$ of the dominating
(zero-vorticity) component in Fig. \ref{fig2}(a), the point of transition
from $d\mu /dN_{-}>0$ to the (presumably) unstable region with $d\mu
/dN_{-}<0$ accurately predicts, for $\gamma =0$, the actual stability
boundary observed in Fig. \ref{stability}. More detailed analysis of this
issue, to be based on computation of eigenvalue spectra for eigenmodes of
small perturbations, will be a subject of a separate work.

\section{Oscillations}

\label{SEC5}

In this section, we address oscillatory dynamics of stable SV solitons. We
start the analysis with a soliton taken in the stability zone ($\gamma =0$, $%
\mu =-0.80$). The oscillations are initiated by a sudden change of the value
of the Zeeman parameter, from $\Omega =1$ fixed above to $\tilde{\Omega}%
=1+\Delta \Omega $ (the added detuning may be both positive and negative, $%
\Delta \Omega \gtrless 0$). The simulations were carried out up to time $%
t=1000$. Figure \ref{fig6}(a) shows the basic frequency of the internal
oscillations in the solitons excited by this sudden perturbation, as a
function of detuning $\Delta \Omega $.

To analyze the excited oscillations, we recorded the largest amplitude of
the dominant zero-vorticity component of spinor wave function, thus creating
a respective time series, $\left( A_{-}\right) _{\max }(t)$. Next, by
applying the Fourier transform, we extract the basic frequency of the
signals. Figure \ref{fig6}(a) demonstrated a nearly linear dependence of the
basic frequency on the detuning, in the considered range of $\Delta \Omega $
(the inset in the figure displays the same result in terms of the dependence
of the corresponding oscillation period on $\Delta \Omega $). Further, Fig. %
\ref{fig6}(b) displays the computed largest, smallest, and average values of
the amplitudes of the vortical and zero-vorticity components, $A_{+}(t)$ and
$A_{-}(t)$, vs. $\Delta \Omega $. Note that the difference between the
largest and smallest values increases with $\Delta \Omega $, being greater
when $\Delta \Omega >0$. For both components, the average values increase
monotonously, indicating that the SV\ become narrower and taller with the
increase of $\Omega $.

To complete the analysis, we consider the time series of the $%
N_{-}(t)/N_{+}(t)$ ratio, which is a record of the relative weight of the
components. Figure \ref{fig6}(c) shows that the highest average value of $%
N_{-}/N_{+}$ is reached at the most negative detuning considered, \textit{viz%
}., $\Delta \Omega =-0.3$, whereas it decreases and stabilizes at $\Delta
\Omega >0$. This fact demonstrates that the decrease in $\Omega $ (which
leads to the contraction of the bandgap (\ref{gap}) populated by the gap
solitons) favors the zero-vorticity component. Furthermore, Fig. \ref{fig6}%
(c) confirms that the internal oscillations of the soliton include the
transfer of atoms between the two spin states. Thus, we infer that the
stable SV solitons remain robust self-trapped modes even in the strongly
excited state.

\section{Conclusion}

\label{SEC6}

In the present work, we have shown the existence of semivortex solitons in
the 2D fermionic spinor field, which includes the Rashba-type spin-orbit
coupling and ZS\ (Zeeman splitting) between the two components, but does not
include the usual kinetic energy (the approximation of \textquotedblleft
heavy atoms"). The spectrum of the system features a bandgap, which may be
populated by gap solitons. The dominant nonlinearity in the system is
provided by the Pauli repulsion, with power $7/3$, as produced by the known
density-functional approximation; the cubic repulsive interaction between
the components is included too, in the general case. We have constructed a
family of gap solitons of the SV\ (semivortex) type, with vortical and
zero-vorticity components in the components with higher and lower Zeeman
energies, respectively. The stability of the SV soliton family has been
identified by means of systematic simulations of the perturbed evolution.
The so identified stability region agrees with the known
anti-Vakhitov-Kolokolov criterion. We have also investigated internal
oscillatory dynamics of SV solitons initiated by a sudden change of the ZS
coefficient. The stable solitons feature robust internal oscillations.

As an extension of the analysis, it is relevant to study the form and
stability of moving SV solitons. This problem is nontrivial, as the
underlying system of Eqs. (\ref{equation1}) and (\ref{equation2}) has no
Galilean invariance. It is known that, if the equations are rewritten in the
reference frame moving with speed $c$, the bandgap (\ref{gap}) shrinks to $%
\mu ^{2}<\sqrt{1-c^{2}}\Omega ^{2}$ \cite{Sakaguchi2018}. Once stable moving
solitons can be found, it may also be interesting to simulate collisions
between ones traveling with opposite speeds $\pm c$.


\section*{Acknowledgments}

PD, LMP, and DL acknowledge partial financial support from FONDECYT 1231020.
JB receives partial financial support through the project with reference:
PID2020-116927RB-C22 of the Ministry of Economy and Competitivity (Spain).
The work of BAM is supported, in part, by grant No. 1695/22 from the Israel
Science Foundation. LMP acknowledges partial financial support from ANID through Convocatoria Nacional Subvenci\'on a Instalaci\'on en la Academia Convocatoria A\~{n}o 2021, Grant SA77210040.


\bibliographystyle{acm}

\bibliography{biblio}

\end{document}